\documentstyle[12pt,times]{article}

\newcommand{\beq}{\begin{equation}}
\newcommand{\eeq}{\end{equation}}
\newcommand{\beqa}{\begin{eqnarray}}
\newcommand{\eeqa}{\end{eqnarray}}
\newcommand{\noi}{\noindent}

\newcommand{\lsim}{\mathrel{\lower4pt\hbox{$\sim$}}
\hskip-12.5pt\raise1.6pt\hbox{$<$}\;}

\newcommand{\gsim}{\mathrel{\lower4pt\hbox{$\sim$}}
\hskip-12.5pt\raise1.6pt\hbox{$>$}\;}

\newcommand{\amzms}{\alpha(m_Z)_{\overline{MS}}}
\newcommand{\aomzms}{\alpha^{-1}(m_Z)_{\overline{MS}}}
\newcommand{\mzms}{(m_Z)_{\overline{MS}}}
\newcommand{\ssthw}{\sin^2\theta_W}
\newcommand{\ssthwz}{\sin^2\theta^0_W}
\newcommand{\ssthweff}{\sin^2\theta^{\rm eff}_W}

\begin{document}

\title{Precision Electroweak Measurements and ``New Physics''}

\author{William J. Marciano\thanks{This manuscript has been authored
under contract number DE-AC02-98CH10886 with the U.S. Department of
Energy. Accordingly, the U.S. Government retains a non-exclusive,
royalty-free license to publish or reproduce the published form of this
contribution, or allow others to do so, for U.S. Government purposes.}\\ 
Brookhaven National Laboratory \\
Upton, New York\ \ 11973 \\}

\maketitle
\begin{abstract}%
\baselineskip 16pt 
 The status of several precisely measured electroweak parameters is
 reviewed. Natural relations among them are shown to constrain the
 Higgs mass, $m_H$, as well as various ``New Physics'' effects.
 Indications of an anomalous $Zb\bar b$ coupling are discussed.
 Constraints on excited $W^\ast$ bosons are given.
\end{abstract}

\section{Fundamental Parameters and Natural Relations}

The SU(2)$_L\times{}$U(1)$_Y$ electroweak sector of the standard model
contains 17 or more fundamental parameters. They include gauge and
Higgs field couplings as well as fermion masses and mixing angles. In
terms of those parameters, predictions can be made with high accuracy
for essentially any electroweak observable. Very precise measurements
of those quantities can then be used to test the standard model, even
at the quantum loop level, or search for small deviations from
expectations which would indicate ``New Physics''.

Some fundamental electroweak parameters have been determined with
extraordinary precision. Foremost in that category is the fine
structure constant $\alpha$. It can best be obtained by comparing the
measured\cite{prl5926} anomalous magnetic moment of the electron,
$a_e\equiv (g_e-2)/2$

\beq
a^{\rm exp}_e = 1159652188 (3) \times 10^{-12} \label{eq1}
\eeq

\noi with the calculated 4 loop QED prediction\cite{hepph9810512}

\beqa
a^{\rm th}_e & = & \frac{\alpha}{2\pi} - 0.328478444
\left(\frac{\alpha}{\pi}\right)^2 + 1.181234 \left(\frac{\alpha}{\pi}
\right)^3 - 1.5098 \left(\frac{\alpha}{\pi}\right)^4 \nonumber \\
& & \quad +1.66\times 10^{-12} \label{eq2}
\eeqa

\noi where the $1.66\times10^{-12}$ comes from small hadronic and weak
loop effects. Assuming no significant ``new physics'' contributions to
$a^{\rm th}_e$, it can be equated with (\ref{eq1}) to give

\beq
\alpha^{-1} = 137.03599959 (40) \label{eq3}
\eeq

\noi That precision is already quite extraordinary. Further improvement
by a factor of 10 appears to be technically feasible and should
certainly be undertaken. However, at this time such improvement would
not further our ability to test QED\null. QED tests require comparable
measurements of $\alpha$ in other processes. Agreement between two
distinct $\alpha$ determinations tests QED and probes for ``new
physics'' effects. After $a_e$, the next best (direct) measurement of
$\alpha$ comes from the quantum Hall effect

\beq
\alpha^{-1} (qH) = 137.03600370 (270) \label{eq4}
\eeq

\noi which is not nearly as precise. Nevertheless, the agreement of
(\ref{eq3}) and (\ref{eq4}) (at the 1.50 level) is a major triumph for
QED up to the 4 loop quantum level.

In terms of probing ``new physics'', one can search for a shift in
$a_e$ by $m^2_e/\Lambda^2_e$ where $\Lambda_e$ is the approximate scale
of some generic new short-distance effect. Current comparison of
$a_e\to \alpha$ and $\alpha(qH)$ explores $\Lambda_e\lsim 100$
GeV\null. To probe the much more interesting $\Lambda_e\sim{\cal O}$
(TeV) region would require an order of magnitude improvement in $a_e$
and about two orders of magnitude error reduction in some direct
precision determination of $\alpha$ 
such as the quantum Hall effect. Perhaps the most likely possibility is
to use the already very precisely measured Rydberg constant in
conjunction with a much improved $m_e$ determination to obtain an
independent $\alpha$.

The usual fine structure constant, $\alpha$, is defined at zero
momentum transfer as is appropriate for low energy atomic physics
phenomena. However, that definition is not well suited for
short-distance electroweak effects. Vacuum polarization loops
screen charges such that the effective (running) electric charge increases at
short-distances. One can incorporate those quantum loop contributions
into a short-distance\cite{prd20274} $\alpha(m_Z)$ defined at
$q^2=m^2_Z$. The main effect comes from lepton loops, which can be very
precisely calculated, and somewhat smaller hadronic loops. The latter
are not as
theoretically clean and must be obtained by combining perturbative
calculations with results of a dispersion relation which employs
${\cal O}(e^+e^-\to{}$hadrons) data. A recent study by Davier and
H\"ocker finds\cite{plb439427}

\beq
\alpha^{-1}(m_Z) = 128.933 (21) \label{eq5}
\eeq

\noi where the uncertainty stems from low energy hadronic loops. Although not
nearly as precise as $\alpha^{-1}$, the uncertainty quoted in
(\ref{eq5}) is impressively small and a tribute to the effort that has
gone into reducing it. (When I first studied this issue in 1979, I
estimated\cite{prd20274} $\alpha^{-1}(m_Z)\simeq 128.5\pm1.0$.)
However, the error in (\ref{eq5}) is still somewhat controversial,
primarily because of its reliance on perturbative QCD down to very low
energies. For comparison, an earlier study by Eidelman and
Jegerlehner,\cite{zpc67585} which relied less on perturbative QCD and
more on $e^+e^-$ data found

\beq
\alpha^{-1}(m_Z) = 128.896 (90) \qquad ({\rm E~\&~J~1995}) \label{eq6}
\eeq

\noi That estimated uncertainty is often cited as more conservative and
therefore employed in $m_H$ and ``new physics'' constraints. As we shall see, the
smaller uncertainty in (\ref{eq5}) has very important consequences for
predicting the Higgs mass. I note that a more recent study\cite{jeger}
by Eidelman and Jegerlehner finds

\beq
\alpha^{-1}(m_Z) = 128.913 (35) \qquad ({\rm E~\&~J~1998}) \label{eq7}
\eeq

\noi which is in good accord with (\ref{eq5}) and also exhibits
relatively small uncertainty. In my subsequent
discussion, I employ the result in (\ref{eq5}), but caution the reader
that a more conservative approach would expand the uncertainty, perhaps
even by as much as a factor of 4 or 5.

A related short-distance coupling, $\amzms$, can be defined by modified
minimal subtraction at scale $\mu=m_Z$. It is particularly useful for
studies of coupling unification in grand unified theories (GUTS) where
a uniform comparitive definition $(\overline{MS})$ of all couplings is called
for.\cite{prl46163} The quantities $\alpha(m_Z)$ and $\amzms$ differ by
a constant, such that\cite{hepph9803453}

\beq
\aomzms = \alpha^{-1}(m_Z) - 0.982 = 127.951 (21) \label{eq8}
\eeq

In weak interaction physics, the most precisely determined parameter is
the Fermi constant, $G_\mu$, as obtained from the muon lifetime. One
extracts that quantity by comparing the experimental value 

\beq
\tau_\mu = 2.197035 (40) \times10^{-6} s \label{eq9}
\eeq

\noi with the theoretical prediction

\beqa
\tau^{-1}_\mu = \Gamma(\mu\to{\rm all}) & = & \frac{G^2_\mu
m^5_\mu}{192\pi^3} f \left(\frac{m^2_e}{m^2_\mu}\right) (1+{\rm R.C.})
\left(1+ \frac{3}{5}\, \frac{m^2_\mu}{m^2_W}\right) \nonumber \\
f(x) & = & 1-8x+8x^3 -x^4 -12x^2\ell nx \label{eq10} 
\eeqa

\noi In that expression R.C. stands for Radiative Corrections. Those
terms are somewhat arbitrary in the standard model. The point being that
$G_\mu$ is a renormalized parameter which is used to absorb most loop
corrections to muon decay. Those corrections not absorbed into $G_\mu$
are explicitly factored out in R.C\null. For historical reasons and in
the spirit of effective field theory approaches, R.C. has been chosen
to be QED corrections to the old V-A four fermion description of  muon
decay.\cite{ap2020} That definition is practical, since the QED
corrections to muon decay in the old V-A theory are finite to all
orders in perturbation theory. In that way, one finds

{\footnotesize
\beq
{\rm R.C.} = \frac{\alpha}{2\pi} \left(\frac{25}{4}-\pi^2\right) \left(
1+\frac{\alpha}{\pi} \left(\frac{2}{3}\ell n \frac{m_\mu}{m_e} -3.7
\right) + \left(\frac{\alpha}{\pi}\right)^2 \left(\frac{4}{9} \ell n^2
\frac{m_\mu}{m_e} - 2.0 \ell n \frac{m_\mu}{m_e} +C\right)\cdots\right)
\label{eq11} 
\eeq}

\noi The leading ${\cal O}(\alpha)$ terms in that expression have been known for
a long time from the pioneering work of Kinoshita and
Sirlin\cite{pr1131652} and Berman.\cite{pr112267} Coefficients of the
higher order logs can be obtained from the renormalization group
constraint\cite{npb29296} 

\beqa
&& \left(m_e \frac{\partial}{\partial m_e} + \beta(\alpha)
\frac{\partial}{\partial\alpha} \right) {\rm R.C.} =0 \nonumber \\
&& \quad \beta(\alpha) = \frac{2}{3}\,\frac{\alpha^2}{\pi} +
\frac{1}{2}\, \frac{\alpha^3}{\pi^2} \cdots \label{eq12}
\eeqa

\noi The -3.7 two loop constant in parenthesis was very recently computed
by van Ritbergen and Stuart.\cite{prl82488} It almost exactly cancels
the leading log two loop correction obtained from the renormalization group
approach (or mass singularities argument) of Roos and
Sirlin.\cite{npb29296} Hence, the original ${\cal O}(\alpha)$ correction in
(\ref{eq9}) is a much better approximation than one might have
guessed. Comparing (\ref{eq9}) and (\ref{eq10}), one finds

\beq
G_\mu = 1.16637 (1)\times10^{-5}{\rm~GeV}^{-2} \label{eq13}
\eeq

There have been several experimental proposals to reduce the
uncertainty in $\tau_\mu$ and $G_\mu$ by a factor of 10. Such
improvement appears technically feasible and, given the fundamental
nature of $G_\mu$, should certainly be undertaken. However, from the
point of view of testing the standard model, the situation is similar
to $\alpha$. $G_\mu$ is already much better known than the other parameters
it can be compared with; so, significant improvement must be made in
other quantities before a more precise $G_\mu$ is required. This point
should become clearer subsequently when I describe other indirect Fermi
constant determinations and their uncertainty (about 100 times worse
than (\ref{eq13})).

Let me emphasize the fact that lots of interesting loop effects have
been absorbed into the renormalization of
$g^2_{2_0}/4\sqrt{2}m^{0^2}_W$ which we call $G_\mu$. Included are top
quark\cite{npb12389} and Higgs loop corrections\cite{npb84132} to the
$W$ boson propagator as well as potential ``new physics'' from SUSY
loops, Technicolor etc. Even tree level effects of possible more
massive gauge bosons such as $W^{\ast^\pm}$ bosons are effectively
incorporated into $G_\mu$. To uncover those contributions requires
comparison of $G_\mu$ with other precisely measured electroweak
parameters which have different quantum loop (or tree level)
dependences. Of course, those quantities must be related to $G_\mu$ in
such a way that short-distance divergences cancel in the comparison.

Fortunately, due to an underlying global SU(2)$_V$ symmetry in the
standard model, there exist natural relations among various bare
parameters\cite{nc16a423} 

\beq
\ssthwz = \frac{e^2_0}{g^2_{2_0}} = 1- (m^0_W/m^0_Z)^2
\label{eq14}
\eeq

\noi Each of those bare unrenormalized  expressions contains
short-distance infinities, but the divergences  are the same.
Therefore, those 
relations continue to hold for renormalized quantities, up to finite,
calculable radiative corrections.\cite{nc16a423} The residual
radiative corrections contain very interesting effects such as $m_t$
and $m_H$ dependence as well as possible ``new physics''. So, for
example, one can relate

\beq
G_\mu = \frac{\pi\alpha}{\sqrt{2} m^2_W (1-m^2_W/m^2_Z)} (1+
rad.~corr.) \label{eq15}
\eeq

\noi and test the predicted radiative corrections, if $m_Z$ and $m_W$
are also precisely known.

Gauge boson masses are not as well determined as $G_\mu$, but they have
reached  high levels of precision. In particular, the $Z$ mass has
been measured with high statistics Breit-Wigner fits to the $Z$
resonance at LEP with the result

\beq
m_Z = 91.1867 (21)~{\rm GeV} \label{eq16}
\eeq

\noi That determination is so good that one must be very precise regarding
the definition of $m_Z$. (Remember the $Z$ has a relatively large width
$\sim2.5$ GeV\null.) The quantity in (\ref{eq16}) is related to the
real part of the $Z$ propagator pole, $m_Z$ (pole), and full width,
$\Gamma_Z$, by\cite{prl672127}

\beq 
m^2_Z = m^2_Z({\rm pole}) + \Gamma^2_Z \label{eq17}
\eeq

\noi The two mass definitions $m_Z$ and $m_Z$ (pole) differ by about 34
MeV, which is much larger than the uncertainty in (\ref{eq16}). Hence,
one must specify which definition is being employed in precision
studies. I note, that the $m_Z$ in (\ref{eq16}) is also more appropriate for
use in low energy neutral current amplitudes.

In the case of the $W^\pm$ bosons, the renormalized mass, $m_W$, is
similarly defined by 

\beq
m^2_W = m^2_W ({\rm pole}) + \Gamma^2_W \label{eq18}
\eeq

\noi That quantity is obtained from studies at $p\bar p$ colliders,
$m_W=80.41 (9)$ GeV, as well as $e^+e^-\to W^+W^-$ at LEPII, $m_W=80.37
(9)$ GeV\null. Together they give

\beq
m_W=80.39 (6)~{\rm GeV}\;. \label{eq19}
\eeq

\noi The current level of uncertainty, $\pm60$ MeV, is large compared
to $\Delta m_Z$. It is expected that continuing efforts at LEPII and
Run II at Fermilab's Tevatron should reduce that error to about $\pm30$
MeV\null. A challenging but worthwhile goal for future high energy
facilities would be to push $\Delta m_W$ to $\pm10$ MeV or better. At
that level, all sorts of interesting ``new physics'' effects are probed (as I
later illustrate). I  note that the $m_W$ defined in (\ref{eq19})
is also the appropriate quantity for low energy amplitudes such as muon
decay.

Another important quantity for precision standard model tests is $m_t$, the
top quark mass. Measurements from CDF and D$\emptyset$ at Fermilab give

\beq
m_t ({\rm pole}) = 174.3\pm5.1~{\rm GeV} \label{eq20}
\eeq

\noi Reducing that uncertainty further is important as we shall subsequently
see. Future Tevatron efforts are expected to reduce the uncertainty in
$m_t$ to about $\pm2$ GeV\null. LHC and NLC studies should bring it
well below $\pm1$ GeV\null.

In addition to masses, the renormalized weak mixing angle plays a
central role in tests of the standard model. That parameter can be
defined in a variety of ways, each of which has its own advocates. I
list three popular examples

\beqa
&\ssthw\mzms & \qquad (\overline{MS} {\rm ~definition~at~}\mu=m_Z)
\qquad\qquad\qquad\, (a) \nonumber \\
&\ssthweff &  \qquad (Z\mu\bar\mu{\rm ~vertex})
\qquad\qquad\qquad\qquad\qquad\qquad (b) \label{eq21} \\
& \ssthw\equiv &\!\!\!\!\!\!\! 1-m^2_W/m^2_Z
\qquad\qquad\qquad \qquad\quad\qquad\qquad\qquad\quad\!\!\!
(c) \nonumber
\eeqa

\noi They differ by finite ${\cal O}(\alpha)$ loop corrections. The
$\overline{MS}$ definition is particularly simple, being defined as the
ratio of two $\overline{MS}$ couplings $\ssthw\mzms\equiv
e^2\mzms /$ $g^2_2 \mzms$. It was introduced for GUT
studies,\cite{prl46163} but is useful for most electroweak analyses.
The effective, $\ssthweff$, weak angle was invented for
$Z$ pole analyses. Roughly speaking, it is defined by the ratio of
vector and axial-vector components (including loops) for the
on-mass-shell $Z\mu\bar\mu$ vertex${}\to1$--$4\ssthweff$. Although
conceptually rather  simple, analytic electroweak 
radiative corrections expressed in terms of $\ssthweff$
are complicated and ugly. Numerically, it is close to the
$\overline{MS}$ definition\cite{pr491160}

\beq
\ssthweff = \ssthw\mzms + 0.00028 \label{eq22}
\eeq

\noi but the analytic structure of the difference is quite complicated. For
those intent on employing $\ssthweff$, a strategy might
be to calculate radiative corrections in terms of $\ssthw$ $\mzms$
and then translate to $\ssthweff$ via (\ref{eq22}). But
why not simply use $\ssthw$ $\mzms$?

Currently, $Z$ pole studies at LEP and SLAC give 

\beqa 
\ssthw\mzms & = & 0.23100\pm 0.00022 \nonumber \\
\ssthweff & = & 0.23128 \pm 0.00022 \label{eq23}
\eeqa

\noi That result includes  measurements of the
left-right asymmetry, $A_{LR}$, at SLAC as well as the various lepton
asymmetries at LEP and SLAC\null. The $A_{LR}$ contribution had for
some time given a relatively low value for the weak mixing angle, but
as statistics have increased it has moved pretty much in line with (\ref{eq23}).
Currently, the $Z\to b\bar b$ forward-backward asymmetries at LEP give a
higher $\ssthweff$ and would bring up the average, if included.
However, the $Zb\bar b$ coupling appears to be
somewhat anomalous; so, one should be cautious when including such
results. I return to this problem later.

Future higher statistics running at SLAC could reduce the uncertainty
in $\ssthweff$ below $\pm0.0002$, mainly from improvements in $A_{LR}$. There
are very good reasons to do even better. One could imagine redoing
$A_{LR}$ at a future polarized lepton-lepton ($e^+e^-$ or $\mu^+\mu^-$) collider,
but with very high statistics. In principle, one might reduce
the uncertainty in $\ssthweff$ to $\pm0.00004$ or lower, an incredible achievement if
possible.

The so-called on-shell or mass definition\cite{prd22971} in
(\ref{eq21}c) also has its advocates. It can be directly obtained from $m_W$ and
$m_Z$ determinations. Indeed, at hadron colliders, the ratio $m_W/m_Z$
can have reduced systematic uncertainties. One could imagine that the
current uncertainty in 

\beq
\ssthw = 1-m^2_W/m^2_Z = 0.2228\pm0.0012 \label{eq24}
\eeq

\noi might be reduced by a factor of about 4 at the LHC\null. Such a
reduction is extremely interesting since the comparison of $\ssthw$ and
$\ssthw\mzms$ provides a clean probe of ``new physics''. It is also
possible (because of a subtle cancellation of certain loop
effects\cite{npb189442}) to measure $\ssthw$ in deep-inelastic $\nu_\mu
N$ scattering. Indeed, a recent Fermilab experiment
found\cite{epjc1509}

\beq
\ssthw = 0.2253\pm0.0019\pm0.0010 \label{eq25}
\eeq

\noi where the first error is statistical and the second systematic.
That single measurement is quite competitive with (\ref{eq24}) and
complements it nicely. One might imagine a future high statistics
effort significantly reducing the error in (\ref{eq25}), but that would
require a new high energy neutrino beam.

All of the above precision measurements can be collectively used to
test the standard model, predict the Higgs mass, and search for ``new
physics'' effects. That ability stems from the natural relations in
(\ref{eq14}) and calculations\cite{prd22971,prd222695} of the radiative
corrections to them. Parametrizing those radiative corrections by
$\Delta r$, $\Delta r\mzms$, and $\Delta\hat r$, one
finds\cite{npb35149}

\beqa
\frac{\pi\alpha}{\sqrt{2}G_\mu m^2_W} & = &
\left(1-\frac{m^2_W}{m^2_Z}\right) (1-\Delta r)
\qquad\qquad\qquad\qquad \qquad\qquad\qquad\! (a)\nonumber \\
\frac{\pi\alpha}{\sqrt{2}G_\mu m^2_W} & = & \ssthw\mzms (1-\Delta
r\mzms) \qquad\qquad\qquad\qquad\quad (b) \label{eq26} \\
\frac{4\pi\alpha}{\sqrt{2}G_\mu m^2_Z} & = & \sin^22\theta_W\mzms
(1-\Delta\hat r) \qquad\qquad\qquad\qquad\qquad\qquad (c) \nonumber
\eeqa

\noi Those expressions contain all one loop corrections to $\alpha$,
muon decay, $m_W$, $m_Z$ and $\ssthw\mzms$ and incorporate some leading two loop
contributions. The quantities $\Delta r$ and $\Delta\hat r$ are
particularly interesting because of their dependence on $m_t$ and
$m_H$. In addition, all three quantities provide probes of ``new
physics''.

Numerically, all three radiative corrections in (\ref{eq26}) contain a
significant contribution from vacuum polarization
effects\cite{prd20274} in $\alpha$, 
about $+7\%$. They are basically the same as the corrections that enter
into the evolution of $\alpha$ to $\alpha(m_Z)$. Leptonic loops
contribute a significant part of that effect and can be very accurately
computed. Hadronic loops are less clean theoretically and lead to a
common uncertainty in $\Delta r$, $\Delta r \mzms$, and $\Delta\hat r$
of

\beq
-\alpha \Delta\alpha^{-1}(m_Z) \label{eq27}
\eeq

\noi For $\Delta\alpha^{-1}(m_Z) = 0.021$ as in (\ref{eq5}), that
amounts to a rather negligible $\pm0.00015$ error. However, for
$\Delta\alpha^{-1}(m_Z)=\pm0.090$ as in (\ref{eq6}), it increases to
$\pm0.00066$. That large an uncertainty would impact precision tests.
If one wishes to avoid that low energy hadronic loop uncertainty,
dependence on $\alpha$ can be circumvented by considering

\beq
\ssthw\mzms = \left(1-\frac{m^2_W}{m^2_Z}\right) (1-\Delta r + \Delta
r\mzms) \label{eq28}
\eeq

\noi Currently, that comparison is not competitive in constraining
$m_H$. However, future significant improvements in $m_W$ could make it
very interesting.

Using $m_t=174.3\pm5.1$ GeV as input, one can compute the radiative
corrections in (\ref{eq26}) as functions of $m_H$. Those results are
illustrated in table~\ref{tabone}. Note that $\Delta r$ is most
sensitive to changes in $m_H$ but also carries the largest uncertainty
from $\Delta m_t=\pm5.1$ GeV ($\pm0.0020$). Hence, efforts to determine $m_H$
from $m_W$ will require a better measurement of $m_t$. On the other
hand, determining $m_H$ from $\ssthw\mzms$ via $\Delta\hat r$ is less
sensitive to $\Delta m_t$ but more sensitive to $\Delta
\alpha^{-1}(m_Z)$. Those dependences are illustrated by the following
approximate relations\cite{plb418188} obtained from (\ref{eq26}a) and
(\ref{eq26}c) 

{\small
\beqa
m_W\!\! & = &\!\! (80.385\pm0.032 \pm0.003{\rm ~GeV})\left(1- 0.00072\ell n
\left(\frac{m_H}{100{\rm ~GeV}}\right)\right. \nonumber \\
& & \qquad \left.- 1\times10^{-4} \ell n^2
\left(\frac{m_H}{100{\rm ~GeV}}\right) \right) \label{eq29} \\
\ssthw\mzms \!\!& = & \!\!(0.23112\pm0.00016\pm0.00006) \left(1+0.00226 \ell
n\left(\frac{m_H}{100{\rm ~GeV}} \right)\right) \label{eq30}
\eeqa}

\noi where the errors correspond to $\Delta m_t=\pm5.1$ GeV and
$\Delta\alpha^{-1}(m_Z) = \pm0.021$ respectively. Note that increasing
$\Delta\alpha^{-1}(m_Z)$ to $\pm0.090$ would significantly compromise
the utility of $\ssthw\mzms$ for determining $m_H$ but have less of an
impact on $m_W$. Predictions for $m_W$ and $\ssthw\mzms$ are illustrated
in table~\ref{tabtwo} for various $m_H$ values.\cite{plb394188}

\begin{table}[tbh]
\begin{center}
\caption{Values of $\Delta r$, $\Delta r\mzms$, and $\Delta\hat r$ for
various $m_H$. A top quark mass of $174.3\pm5.1$ GeV and
$\alpha^{-1}(m_Z) = 128.933 (21)$ are assumed. \label{tabone}}
\begin{tabular}{lccc}
$m_H$ (GeV) & $\Delta r$ & $\Delta r\mzms$ & $\Delta\hat r$ \\
& $\pm0.0020\pm0.0002$ & $\pm0.0001\pm0.0002$ & $\pm0.0005\pm0.0002$ \\
\\
~75 & 0.03402 & 0.06914 & 0.05897 \\
100 & 0.03497 & 0.06937 & 0.05940 \\
125 & 0.03575 & 0.06955 & 0.05974 \\
150 & 0.03646 & 0.06964 & 0.06000 \\
200 & 0.03759 & 0.06980 & 0.06042 \\
400 & 0.04065 & 0.07005 & 0.06144 \\
\end{tabular}
\end{center}
\end{table}

\begin{table}[tbh]
\begin{center}
\caption{Predictions for $m_W$ and $\ssthw\mzms$ for various $m_H$
values. \label{tabtwo}}
\begin{tabular}{lcc}
$m_H$ (GeV) & $m_W$ (GeV) & $\ssthw\mzms$ \\ \\
~75 & 80.401 & 0.23097 \\
100 & 80.385 & 0.23112 \\
125 & 80.372 & 0.23124 \\
150 & 80.360 & 0.23133 \\
200 & 80.341 & 0.23148 \\
400 & 80.289 & 0.23184 \\
\end{tabular}
\end{center}
\end{table}

Employing $m_W =80.39\pm0.06$ GeV and $\ssthw\mzms=0.23100\pm0.00022$,
one finds

\beqa
m_H & = & 92^{+141+75+5}_{-67-35-5}{\rm ~GeV} \qquad ({\rm from~}m_W)
\label{eq31} \\
m_H & = & 79^{+41+28+9}_{-27-21-8}{\rm ~GeV} \qquad ({\rm
from~}\ssthw\mzms) \label{eq32}
\eeqa

\noi where the second and third errors correspond to $\Delta
m_t=\pm5.1$ GeV and $\Delta\alpha^{-1}(m_Z) = \pm0.021$. Several
features of those predictions are revealing. The first is that
$\ssthw$ $\mzms$ currently gives a very good (best) determination of $m_H$.
Note, however, the uncertainties scale as the central value; so, the
relatively small value, 79 GeV, helps reduce the uncertainties. Also, a
larger $\Delta\alpha^{-1}(m_Z)=\pm0.090$ would significantly increase
the overall uncertainty.\cite{hepph9812332} In the case of $m_W$, one needs a
better measurement of that parameter along with improvement in $m_t$, if
it is to pinpoint $m_H$.

Taken together, (\ref{eq31}) and (\ref{eq32}) are very suggestive of a
relatively light Higgs scalar not far from the current LEPII bound from
non observation of $e^+e^-\to ZH$

\beq
m_H>89.8 {\rm ~GeV} \qquad (\sqrt{s}=183 {\rm ~GeV~data}) \label{eq33}
\eeq

\noi Preliminary studies of $\sqrt{s}=189$ GeV $e^+e^-$ data indicate
that bound will soon rise to $\sim95$ GeV\null. Future upgrades to
$\sqrt{s}\simeq 200$ GeV will push the Higgs discovery potential to
$\sim105$ GeV\null. In addition, searching for the Higgs via associated
$W^\pm H$ and $ZH$ at the Fermilab $p\bar p$ collider during Run II
promises discovery up to $m_H\sim 115$ GeV, perhaps even higher. Higgs
discovery may soon be at hand.

\section{The $Zb\bar b$ Problem}

Currently, the only real anomaly in $Z$ pole measurements seems to
involve $b\bar b$ final states. The LEP $b\bar b$ forward-backward
asymmetry and SLAC $b\bar b$ left-right forward-backward asymmetry are
consistent with a $3\sigma$ deviation for $A_b$ from the standard model
expectation

\beqa
A^{\rm exp}_b/ A^{\rm theory}_b & = & 0.96\pm0.02 \nonumber \\
A_b & = & \frac{g^2_L(b) - g^2_R(b)}{g^2_L(b) + g^2_R(b)} \label{eq34}
\eeqa

\noi where $g_L(b)$ and $g_R(b)$ are the $Z$ couplings to left and
right-handed $b$ quarks, normalized to have  standard model values (at
tree level)

\beqa
g_L(b) & = & -\frac{1}{2} + \frac{1}{3}\ssthw \simeq -0.423 \nonumber
\\
g_R(b) & = & \frac{1}{3}\ssthw \simeq +0.077 \label{eq35}
\eeqa

\noi At the same time, the quantity $R_b\equiv \Gamma(Z\to b\bar
b)/\Gamma(Z\to{}$hadrons) exhibits very good accord with standard model
expectations

\beq
R^{\rm exp}_b/R^{\rm theory}_b = 1.004\pm0.004 \label{eq36}
\eeq

\noi and thereby leads to the constraint

\beq
g^2_L(b) +g^2_R(b) = 0.1858\pm0.0010 \label{eq37}
\eeq

\noi Solving (\ref{eq34}) and (\ref{eq37}) together gives 

\beqa
g^2_L(b) & = & 0.1746\pm0.0020 \nonumber \\
g^2_R(b) & = & 0.0112\pm0.0018 \label{eq38}
\eeqa

\noi In terms of deviations from the standard model, $\delta g_L(b)$
and $\delta g_R(b)$, one finds (ignoring the negative $g_R(b)$ and
positive $g_L(b)$ solutions)

\beqa
\delta g_L(b) & = & 0.005\pm0.002 \nonumber \\
\delta g_R(b) & = & 0.0287\pm0.0088 \label{eq39}
\eeqa

\noi The $\delta g_L(b)$ deviation on its own amounts to only a - 1\%
shift and  could probably be
interpreted as a ``new physics'' quantum loop correction; however, such
a large $\delta g_R(b)$ shift of 40\% is very difficult to explain. For that reason,
most theorists have dismissed the above $3\sigma$ deviation as
experimental in origin, i.e.\ stemming from a statistical or systematic effect,
rather than an indication of ``new physics''. Nevertheless, it is
amusing to contemplate other potential consequences of non-zero
$\delta g_L(b)$ and $\delta g_R(b)$ of the magnitude in (\ref{eq39}).
First, I note that deviations of similar magnitude cannot occur in
$\delta g_L(d)$ and $\delta g_R(d)$; otherwise they would have been
observed in atomic parity violation and $\nu_\mu N$ experiments. Furthermore, it
is unlikely that they are present in $\delta g_R(s)$ and $\delta
g_L(s)$. If that were the case, one would expect (but could avoid)
induced $s\to d$ 
flavor-changing weak neutral currents which could significantly enhance
$K_L\to\mu^+\mu^-$, $K^+\to\pi^+\nu\bar \nu$ etc; and that seems not to
be the case. If one concludes that
the anomaly occurs only in  $Zb\bar b$, it is still likely that related
new flavor
changing $b\to s$, $b\to d$, and $s\to d$ weak neutral currents would
 occur. The predicted magnitude of those effects depends on the degree
 and nature of
quark mixing; however, generically interesting observable consequences
almost certainly  result. It will be interesting to see if anomalies in
$Z\to b\bar b$ 
asymmetries persist as the data is further scrutinized and whether FCNC
$b$ (and $K$) decays will be in accord with Standard Model expectations
or also exhibit anomalies.

\section{Muon Decay and the $S$, $T$, $U$ Parameters}

As previously discussed, muon decay provides a very precise
determination of $G_\mu=1.16637(1)\times10^{-5}$ GeV$^{-2}$ which
contains within it potential ``new physics'' effects. For example,
heavy chiral fermions present in 4th generation models or technicolor
theories would contribute to gauge boson self energies. Those loop effects
would show up in the $\Delta r$, $\Delta r\mzms$ and $\Delta \hat r$ of
(\ref{eq26}) as additional contributions. One way to unveil or
constrain such effects is to define 
Fermi constants in terms of $\alpha$, $\ssthw\mzms$, $m_W$, and $m_Z$
via (\ref{eq26})

\beqa
G^{(1)}_F & = & \frac{\pi\alpha}{\sqrt{2} m^2_W (1-m^2_W/m^2_Z) (1-\Delta
r)} \qquad\qquad\qquad \qquad\qquad\qquad\!\! (a) \nonumber \\
G^{(2)}_F & = & \frac{\pi\alpha}{\sqrt{2}m^2_W\ssthw\mzms (1-\Delta
r\mzms)} \qquad\qquad\qquad\qquad\! (b) \label{eq40} \\
G^{(3)}_F & = & \frac{4\pi\alpha}{\sqrt{2}m^2_Z \sin^22\theta_W\mzms
(1-\Delta \hat r)} \qquad\qquad\qquad \qquad\qquad\quad (c) \nonumber
\eeqa

\noi Comparison of those quantities with $G_\mu$ tests the standard
model and probes for possible ``new physics'' in the $\Delta r$,
$\Delta r\mzms$, and $\Delta\hat r$. If there is no ``new physics'' one
should find $G_\mu=G^{(1)}_F = G^{(2)}_F = G^{(3)}_F$.

To examine the situation, we take $m_H=125$ GeV as our central value
and allow for the range $75<m_H<200$ GeV\null. Then from the values in
table~\ref{tabone} and the measured $m_W$, $m_Z$, $\ssthw\mzms$ and
$\alpha$, one finds

\beqa
G^{(1)}_F & = & 1.1676 (55) \times 10^{-5} {\rm ~GeV}^{-2}
\qquad\qquad\qquad\qquad\qquad\qquad \!(a) \nonumber\\
G^{(2)}_F & = & 1.1671 (21) \times 10^{-5} {\rm ~GeV}^{-2}
\qquad\qquad\qquad\qquad\qquad\qquad (b) \label{eq41} \\
G^{(3)}_F & = & 1.1672 (14) \times 10^{-5} {\rm ~GeV}^{-2}
\qquad\qquad\qquad\qquad\qquad\qquad (c) \nonumber 
\eeqa

\noi where the uncertainties reflect errors in $m_t$, $m_H$,
$\Delta^{-1}\alpha(m_Z)$, $m_W$, and $\ssthw\mzms$. The excellent
agreement between those quantities and $G_\mu=1.16637 (1)\times10^{-5}$
GeV$^{-2}$ obtained from muon decay is quite remarkable. It shows no
indication of ``new phys\-ics''. Note also that the uncertainty in even
the most precise $G^{(3)}_F$ is more than 100 times the current error in
$G_\mu$. Hence, improving $G_\mu$ further would not sharpen such tests,
improving $\ssthw(m_Z)$, $m_W$, $m_t$ and measuring $m_H$ would.

As an example of the utility of (\ref{eq41}), consider the deviations
expected from heavy chiral fermion doublets. The appendage of such
particles to the standard model modifies gauge  boson self-energies.
Those effects shift the radiative corrections in (\ref{eq26}). Such
shifts are conveniently parametrized by the $S$, $T$, and $U$
parameters of Peskin and Takeuchi\cite{prl65964,prl652963}

\beqa
\delta\Delta r & = & 0.0166S-0.0258T-0.0195U \;\quad\qquad\qquad\qquad\qquad (a)
\nonumber \\
\delta \Delta r\mzms & = & 0.0084 (S+U) \qquad\qquad\qquad\qquad\qquad\qquad\qquad\qquad (b)
\label{eq42} \\
\delta\Delta\hat r & = & 0.011S-0.00782T \,\qquad\qquad\qquad\qquad\qquad\qquad\qquad (c)
\nonumber
\eeqa

\noi Hence, if $S$, $T$ and $U\ne0$, one expects the relationships

\beqa
G_\mu & = & G^{(1)}_F(1+0.0166S-0.0258T-0.0195U) \;\quad\qquad\qquad\qquad (a)
\nonumber \\
G_\mu & = & G^{(2)}_F (1+0.0084(S+U))
\qquad\qquad\qquad\qquad\qquad\qquad\qquad (b) \label{eq43} \\
G_\mu & = & G^{(3)}_F (1+0.011S-0.00782T) \,\qquad\qquad\qquad\qquad\qquad\qquad (c)
\nonumber
\eeqa

\noi In technicolor models, one has the generic
prediction\cite{prl65964} $S\sim {\cal O}(+1)$ which would lead to
about a 1\% difference between the $G_\mu$ and $G^{(i)}_F$. However,
(\ref{eq41}) exhibits no such effect. In fact it constrains that
quantity at ${\cal O}(0.1\%)$. A global fit to all electroweak data
(for $m_H\simeq100$ GeV) gives\cite{hepph9809352}
\newpage

\beqa
S & = & -0.17^{+0.17}_{-0.12} \qquad\qquad\qquad\qquad\qquad\qquad\qquad\qquad\qquad\qquad (a)
\nonumber \\
T & = & -0.16^{+0.15}_{-0.18} \,\qquad\qquad\qquad\qquad\qquad\qquad\qquad\qquad\qquad\qquad (b)
\label{eq44} \\
U & = & 0.19\pm0.21 \qquad\qquad\qquad\qquad\qquad\qquad\qquad\qquad\qquad\qquad (c)
\nonumber
\eeqa

\noi Those constraints are consistent with (\ref{eq43}) and
(\ref{eq41}). If one assumes $m_H\sim {\cal O}$(1 TeV) as would be more
appropriate for technicolor, one finds $S=-0.29\pm0.14$ which is even
more incompatible with $S\sim {\cal O}(+1)$.  Therefore, if dynamical
electroweak symmetry breaking is to be consistent with precision
measurements, the dynamics must be very novel to render $S\sim0$.

\section{Extra Dimensions and $W^{\ast^\pm}$ Bosons}

Another interesting ``new physics'' scenario involves excited
$W^{\ast^\pm}$ bosons which may arise in theories with extra compact
dimensions\cite{plb27021} (Kaluza-Klein excitations) or models with
composite gauge bosons. Assuming fermionic couplings to $W^{\ast^\pm}$
indentical to those of $W^\pm$, $g^\ast_2=g_2$, direct searches at the
Tevatron lead to  the bound\cite{pdgtables}

\beq
m_{W^\ast} > 720 {\rm ~GeV} \qquad (95\% {\rm ~CL}) \label{eq45}
\eeq

If such bosons exist, they would also contribute to low energy charged
current processes such as muon decay and be incorporated into $G_\mu$.
They would replace $g^2_2/m^2_W$ in the decay amplitude by
$g^2_2/\langle m^2_W\rangle$ where 

\beq
\frac{1}{\langle m^2_W\rangle} = \frac{1}{m^2_W} +
\frac{(g^\ast_2/g_2)^2}{m^2_{W^\ast}} +
\frac{(g^{\ast\ast}_2/g_2)^2}{m^2_{W^{\ast\ast}}} + \cdots \label{eq46}
\eeq

\noi As long as the signs are positive, the effective low energy mass
$\langle m_W\rangle$ is always smaller than $m_W$, since the reciprocal
sum acts like resistors in parallel. Therefore, if $W^\ast$ bosons
exist, $G_\mu$ should be larger than the $G^{(i)}_F$ in (\ref{eq41}).
However, there is no such indication. Quantitatively, one expects

\beq
G_\mu = G^{(i)}_F \left( 1+ C\left(\frac{g^\ast_2}{g_2}\right)^2
\frac{m^2_W}{m^2_{W^\ast}} \right) \label{eq47}
\eeq

\noi where

\beq
C= 1+ \left(\frac{g^{\ast\ast}_2}{g^\ast_2} \right)^2
\frac{m^2_{W^{\ast}}}{m^2_{W^{\ast\ast}}} + \cdots >1 \label{eq48}
\eeq

\noi In the simplest extra dimension theory, one might typically expect
$C=\sum\limits^\infty_{n=1} 1/n^2 = \pi^2/6$. More realistic scenarios
can lead to even larger $C$. Here, I am interested only in lower bounds
on $m_{W^\ast}$; so, $C$ will not enter as long as $C\ge1$.

Comparing (\ref{eq47}) and (\ref{eq41}b) gives (for $m_H\lsim200$ GeV)

\beq
m_{W^\ast} > 1.67 \left(\frac{g^\ast_2}{g_2}\right) {\rm ~TeV} \qquad
(95\% {\rm ~CL}) \label{eq49}
\eeq

\noi which is very constraining. It suggests that the radius of the
extra dimensions $R\simeq 1/m_{W^\ast} < 1\times10^{-17}
(g_2/g^\ast_2)$ cm.\ and that continuing searches for $W^\ast$ bosons
at the Tevatron are likely to yield null results. Of course, the bound
can be significantly relaxed if $g^\ast_2<<g$.

One can improve the bound in (\ref{eq49}) by employing $G^{(3)}_F$
rather than $G^{(2)}_F$ in comparison with $G_\mu$. One finds

\beq
m_{W^\ast} > 2.27 \left(\frac{g^\ast_2}{g_2}\right) {\rm ~TeV} \qquad
(95\% {\rm ~CL}) \label{eq50}
\eeq

\noi However, that bound is subject to a larger dependence on $m_t$,
$m_H$, and ``new physics'' effects. Further improvements in
$\ssthw\mzms$ and $m_W$ could push the $m_{W^\ast}$ sensitivity to
${\cal O}$(5 TeV) which is competitive with LHC capabilities.

\section{Conclusion}

Precision electroweak measurements have tested the standard model at
the $\pm0.1\%$ level. As a byproduct, they have been used to predict the
large top quark mass and now suggest a relatively light
Higgs\cite{hepph9809352} 

\beq
m_H < 255 {\rm ~GeV} \qquad (95\% {\rm ~CL}) \label{eq51}
\eeq

\noi with values around 100 GeV favored. Discovery of the Higgs scalar
may be close.

The good agreement between theory and experiment severely constrains
the possible ``new physics'' one can append to the standard model. For
example, the $S$ parameter must be near zero. That finding leaves
little room for additional chiral fermion doublets such as a fourth
generation of fermions and requires
dynamical symmetry breaking scenarios to exhibit novel dynamics which
respects that constraint (a difficult task). Other types of ``new
physics'' such as relatively large extra dimensions, SUSY, $Z^\prime$
bosons etc.\ are also being constrained by such measurements. So far,
there are no signs of ``new physics''.
Nevertheless, we must continue to probe shorter distances and search
for new phenomena. Surprises are certainly waiting to be unveiled.

\end{document}